
\input amstex
\documentstyle{fic}
\NoBlackBoxes

\topmatter

\title Black Hole Entropy and Renormalization  \endtitle
\author Jean-Guy Demers, Ren\'e Lafrance and Robert C. Myers \\
\address Department of Physics,  McGill University \\
	Montr\'eal, Qu\'ebec, Canada, H3A 2T8
\email jgdemers@hep.physics.mcgill.ca \endemail
\email lafrance@hep.physics.mcgill.ca  \endemail
\email rcm@hep.physics.mcgill.ca  \endemail
\endaddress
\endauthor

\leftheadtext{J.-G. Demers, R. Lafrance and R.C. Myers}
\rightheadtext{Black Hole Entropy and Renormalization}

\abstract
Using a new regulator, we examine
't Hooft's approach for evaluating black hole entropy through
a statistical-mechanical counting of states for a scalar field
propagating outside the event horizon.
We find that this calculation
yields precisely the one-loop renormalization of the standard
Bekenstein-Hawking formula, $S={\Cal A}/(4G)$.
Thus our result provides evidence confirming
a suggestion by Susskind and Uglum regarding black hole entropy.

\endabstract


\cvolyear {0000\endgraf McGill/95-39 \endgraf gr-qc/9507042 }
\thanks
Talk given by R. Lafrance at the 6 Canadian Conference on General
Relativity and Realtivistic Astrophysics, Frederiction, May 25-27 1995.
 \endgraf
This research was supported by NSERC of Canada and Fonds FCAR du Qu\'ebec.
\endthanks
\endtopmatter

\document

\head 1\enspace Introduction\endhead

It is now over twenty years since Bekenstein introduced the idea
that black holes carry an intrinsic entropy proportional
to the surface area of the event horizon measured in
Planck units, {\it i.e.},\ ${\Cal A}/{\ell_p}^2$
(\cite{Bekenstein [1972]}).
Hawking's discovery (\cite{Hawking [1975]}) that, in quantum field
theory, black holes actually
generate thermal radiation allowed the determination of a precise
formula for this entropy: $S={\Cal A}/(4G)={\Cal A}/(4{\ell_p}^2)$.
(We adopt the standard conventions of setting
$\hbar=c=k_{\scriptscriptstyle B}=1$, but we will explicitly retain
Newton's constant, $G$, in our analysis.)
This Bekenstein-Hawking formula is applicable for any black hole
solution of Einstein's equations. In more general gravitational
theories, black hole entropy is no longer given by
the Bekenstein-Hawking formula
(\cite{Wald [1993]}; \cite{Visser [1993]}; \cite{Jacobson,
Kang, Myers [1994]}; \cite{Iyer, Wald [1994]}).
However, the entropy is given by the integral of
a geometric density over a cross section of the horizon. For
example, in a theory described by the effective gravitational action
$$
I=\int \, d^4x \, \sqrt{-g} \biggl[ \frac{R}{16\pi G} +\frac{\gamma}{4\pi}
	R^{abcd}R_{abcd} \biggr] \ ,
\tag{1.1}
$$
the entropy of a black hole is
$$
S=\oint_H d^2x \, \sqrt{h} \biggl[ \frac{1}{4G} -\frac{\gamma}{4\pi}
	R^{abcd} \hat{\epsilon}_{ab} \hat{\epsilon}_{cd} \biggr] \ .
\tag{1.2}
$$
Here $H$ is a space-like cross section of the horizon, $h_{ab}$ is the
induced metric on $H$, and $\hat{\epsilon}_{ab}$ is the binormal to $H$.

Our present
understanding of black hole entropy is limited to a thermodynamic
framework. Many attempts have been made  to define black hole entropy
within a statistical mechanical context, but
despite a great deal of effort, such a microphysical interpretation
is still lacking. A common feature of many of the attempts
is that they yield a black hole entropy
proportional to the area, but with a divergent coefficient.
A recent suggestion by Susskind and Uglum [1994] is
that these divergences can
be absorbed in the Bekenstein-Hawking formula as a renormalization
of Newton's constant. We investigate this possibility by
using a common regulator for a calculation of the renormalization
of the gravitational coupling constants, and a statistical
calculation of black hole entropy. This allows a precise comparison
of the divergences appearing in these two calculations. Our
results support the suggestion of Susskind and Uglum [1994]. A
more complete description of this work appears in Demers, Lafrance,
Myers [1995a].

\head 2\enspace Renormalization of the gravitational action\endhead

In the study of the one-loop effective action (see {\it e.g.}
\cite{Birrell, Davies [1982]}),
one may start with the gravitational action
$$
I_g=\int d^4x\, \sqrt{-g} \biggl[ -\frac{\Lambda_B}{8\pi G_B} +
\frac{R}{16\pi G_B} +\frac{\alpha_B}{4\pi} R^2 +
\frac{\beta_B}{4\pi} R_{ab}R^{ab} +
\frac{\gamma_B}{4\pi} R_{abcd}R^{abcd} + \ldots\ \bigr]
\tag{2.1}
$$
where $\Lambda_B$ is the cosmological constant, $G_B$ is Newton's constant,
while $\alpha_B$, $\beta_B$ and $\gamma_B$ are dimensionless coupling
constants for the interactions quadratic in the curvature.
The subscript $B$ indicates that these are all bare
coupling constants. The ellipsis indicates
that the action may also include other covariant,
higher-derivative interactions, but no terms beyond those shown
will be of interest in the present analysis. We also include the action for
a minimally coupled neutral scalar field,
$$
I_m=-\frac{1}{2} \int d^4x \, \sqrt{-g} \biggl[ g^{a b} \nabla_{\! a}\phi
 \nabla_{\! b} \phi +m^2 \phi^2 \biggr]\ .
\tag{2.2}
$$
We wish to determine the effective action for the metric which
results when in the path integral the scalar field is integrated out.
This integration is simply gaussian, yielding
the square root of the determinant of the propagator. Thus the
contribution to the effective gravitational
action, which is the logarithm of this result,
is given by  $W(g)=-{i\over2} \roman{Tr}
\log[-G_{ F}(g,m^2)]$ (\cite{Birrell, Davies [1982]}).
Of course, as it stands, this expression is divergent and it must first be
regulated to be properly defined. The divergences of this one-loop
effective action,
as well as its metric dependence, are then easily identified using
an adiabatic expansion for the DeWitt-Schwinger proper time representation
of the propagator. To regulate our calculations,
we choose the Pauli-Villars method. In this scheme,
one introduces fictitious regulator fields with very large masses. Some
of these fields are also quantized with the
``wrong'' statistics. For the present four-dimensional scalar field theory,
one needs five regulator fields, and the total action for the matter fields
becomes
$$
I_m=-\frac{1}{2}\sum_{i=0}^5
 \int d^4x \, \sqrt{-g} \biggl[ g^{a b} \nabla_a \phi_i
 \nabla_b \phi_i +m_i^2 \phi_i^2 \biggr] \ .
\tag{2.3}
$$
Here, $\phi_0=\phi$ is the physical scalar
with mass $m_0=m$,  $\phi_1$ and $\phi_2$
 are two anticommuting fields with mass $m_{1,2}=\sqrt{\mu^2+m^2}$,
$\phi_3$ and $\phi_4$ are two commuting fields with mass $m_{3,4}=
\sqrt{3\mu^2+m^2}$, and $\phi_5$ is an anticommuting
field with mass $m_{5}=\sqrt{4\mu^2+m^2}$. Now, each field makes a
contribution
to the effective action as discussed above, except that as a result of
their anticommuting statistics, the contributions of
$\phi_2,\ \phi_3$ and $\phi_5$ have the opposite sign,
{\it i.e.}, $W(g)\simeq+{i\over2}\roman{Tr}\log[-G_ F(g,m_i^2)]$.
We will focus on the divergent terms of the one-loop effective action:
$$
W_{div} = \frac{1}{32\pi^2}\int d^4x\,\sqrt{-g}\, [ -C\,a_0(x)+B\, a_1(x)
+ A a_2(x) ]\ .
\tag{2.4}
$$
In this expression, the coefficients $a_0$, $a_1$, $a_2$ are functionals
of the local geometry
$$
\aligned
a_0&=1 \; \; \;\;\;\;\;\;\;\;\;\;\;\;\;\;\;   a_1= \frac{1}{6} R \\
a_2&= \frac{1}{180} R^{abcd} R_{abcd} -\frac{1}{180} R^{ab} R_{ab}
+\frac{1}{30} \square R +\frac{1}{72} R^2
\endaligned
\tag{2.5}
$$
and  $A$, $B$ and $C$ are constants
which depend on the masses, $m$ and $\mu$, and which diverge for
$\mu\rightarrow \infty$:
$$
\aligned
A=& \ln \frac{4\mu^2+m^2}{m^2} +2\ln \frac{\mu^2+m^2}{3\mu^2+m^2} \\
B=& \mu^2 \biggl[2 \ln \frac{3\mu^2+m^2}{\mu^2+m^2}
+4\ln \frac{3\mu^2+m^2}{4\mu^2+m^2} \biggr] +
m^2 \biggl[ \ln \frac{m^2}{4\mu^2+m^2} +2\ln \frac{3\mu^2+m^2}
{\mu^2+m^2} \biggr]  \\
C =&   \mu^4 \biggl[
8\ln \frac{3\mu^2+m^2}{4\mu^2+m^2} +\ln \frac{3\mu^2+m^2}
{\mu^2+m^2}\biggr]  +2  m^2\mu^2 \biggl[  \ln \frac{3\mu^2+m^2}
{\mu^2+m^2} +2\ln \frac{3\mu^2+m^2}{4\mu^2+m^2} \biggr]  \\
 &\qquad +\frac{m^4}{2} \biggl[ \ln \frac{m^2}{4\mu^2+m^2} +2\ln
\frac{3\mu^2+m^2}{\mu^2+m^2} \biggr]  \ .
\endaligned
\tag{2.6}
$$
For large values of $\mu$, the constants $A,\ B$ and
$C$ grow to leading order as $\ln(\mu/m),\ \mu^2$ and $\mu^4$,
respectively, but they also contain subleading and finite
contributions.
Combining the scalar one-loop action with the original bare action in
eq.~\rom{(2.1)}, we can identify the renormalized coupling
constants in the effective gravitational action
$$
\aligned
I_{eff} & =I_g\ +\ W   \\
& =\int d^4 x \sqrt{-g} \biggl[   -\frac{1}{8\pi} \biggl(
\frac{\Lambda_B}{G_B} +\frac{C}{4\pi} \biggr) +\frac{R}{16\pi} \biggl(
\frac{1}{G_B} +\frac{B}{12\pi} \biggr) +\frac{R^2}{4\pi} \biggl(
\alpha_B +\frac{A}{576\pi} \biggr)  \\
&\qquad\qquad
 +\frac{1}{4\pi}R_{ab}R^{ab} \biggl( \beta_B-\frac{A}{1440\pi}\biggr)
+\frac{1}{4\pi}R_{abcd}R^{abcd} \biggl(\gamma_B +\frac{A}{1440\pi}\biggr)
+\ldots \biggr]   \\
\endaligned
\tag{2.7}
$$
where in the action we discard the total derivative term $\square R$
occurring in $a_2$. In particular from eq.~\rom{(2.7)},
we obtain the renormalized Newton's constant:
$$
\frac{1}{G_R}=\frac{1}{G_B}+\frac{B}{12\pi}\ .
\tag{2.8}
$$
In eq.~\rom{(2.7)}, divergent renormalizations also occur for
the cosmological constant $\Lambda_B$ and the quadratic-curvature
coupling constants $\alpha_B$, $\beta_B$ and $\gamma_B$.
The higher order bare coupling
constants (beyond those explicitly shown) would receive finite
renormalizations but they will play no role in the present analysis.

\head 3\enspace Statistical entropy \endhead

't Hooft's statistical mechanical calculation of black hole entropy
(\cite{'t Hooft [1985]}) involves counting the states for
a thermal ensemble of a scalar field propagating just outside a
black hole. Following 't Hooft's calculation,
we introduce a fixed background metric, which is
a Schwarzchild black hole
$$
ds^2=-\biggl( 1-\frac{r_0}{r} \biggr) dt^2 + \biggl( 1-\frac{r_0}{r}
\biggr)^{\!-1} dr^2 +r^2 (d\theta^2 +\sin^2 \theta \, d\varphi^2 ) \ .
\tag{3.1}
$$
In this background, we consider a minimally coupled scalar
field satisfying the Klein-Gordon equation
$$
(\square -m^2 ) \phi(x) =0 \ .
\tag{3.2}
$$
Because of the infinite blue-shift at the horizon, 't Hooft introduced a
``brick wall''
cut-off: $\phi(x)=0$ for $r \leq r_0+h$, with $h \ll r_0$. To eliminate
infrared divergences, a second cut-off is introduced at a large radius
$L \gg r_0$: $\phi(x)=0$ for $r \geq L$.

To calculate the entropy, we consider the free energy of a thermal
ensemble of scalar particles at inverse temperature $\beta$
$$
\beta F=\int_0^\infty dE \, \frac{dg}{dE} \, \ln (1-e^{-\beta E}) \ ,
\tag{3.3}
$$
where $dg(E)/dE$ is the density of modes with energy $E$. In the
WKB appproximation, the number of states $g(E)$ is given by
$$
g(E)=\frac{2}{3\pi} \int_{r_0+h}^{L} dr \, r^2 \biggl( 1-\frac{r_0}{r}
\biggr)^{\!-2} \biggl[ E^2- \biggl( 1-\frac{r_0}{r} \biggr) m^2
\biggr]^{3/2} \ .
\tag{3.4}
$$
{}From eq.~(3.4), it is apparent that
$F$ will diverge if one takes the limit $h \rightarrow 0$.
However, the divergences are removed if we add the same Pauli-Villars fields
that were introduced in section~2. With the regulator fields and
$h\rightarrow 0$, the free energy becomes
$$
\bar{F}= -\frac{2}{3\pi} \int_0^\infty \frac{dE}{e^{\beta E}-1}
\int_{0}^{L} dr \, r^2 \biggl( 1-\frac{r_0}{r}
\biggr)^{\!-2} \sum_{i=0}^5 \Delta_i
\biggl[ E^2- \biggl( 1-\frac{r_0}{r} \biggr) m_i^2
\biggr]^{3/2}  \ ,
\tag{3.5}
$$
where $\Delta_0=\Delta_3=\Delta_4=1$ for the commuting fields and $\Delta_1=
\Delta_2=\Delta_5=-1$ for the anticommuting fields. Now the
divergent contributions at the horizon are:
$$
\bar{F} =-r_0^3 \biggl[ \frac{\pi}{6\beta^2} B +\frac{8\pi^3}{45 \beta^4} A
\biggr] \ ,
\tag{3.6}
$$
where $A$ and $B$ are the same constants
given by Eq. (2.6).

The corresponding entropy is then:
$$
S_q=\beta^2 \frac{\partial F}{\partial \beta} \biggl|_{\beta=4\pi r_0}
= \frac{{\Cal A}}{4} \frac{B}{12 \pi} +\frac{A}{90} \ .
\tag{3.7}
$$
Using eq.~\rom{(1.2)}, the bare entropy for the
Schwarzchild black hole
related to the gravitational action \rom{(2.1)} is
$$
S_B=\frac{\Cal A}{4G_B} +16 \pi \gamma_B \ .
\tag{3.8}
$$
Adding this result to $S_q$ yields:
$$
\aligned
S=S_B+S_q &=\frac{{\Cal A}}{4} \biggr( \frac{1}{G_B} +\frac{B}{12\pi}
\biggl)
+16\pi \biggr( \gamma_B +\frac{A}{1440} \biggl) \\
&= \frac{{\Cal A}}{4G_R} +16 \pi \gamma_R \ .\\
\endaligned
\tag{3.9}
$$
Thus this final entropy expression contains precisely the renormalized
coupling constants appearing in eq.~\rom{(2.7)}.

\head 3 \enspace Conclusion  \endhead

Through the use of a Pauli-Villars regulator, we have
found evidence for the suggestion by
Susskind and Uglum [1994] --- {\it i.e.,}
the divergences appearing in 't~Hooft's statistical mechanical
calculation of black hole entropy can be absorbed
by a renormalization of the gravitational coupling constants.
(Similar results were found by
\cite{Fursaev, Solodukhin [1994] and Solodukhin [1995]}.)
We have extended these calculations to a Reissner-Nordstr\"om background,
and the same conclusions follow
(\cite{Demers, Lafrance, Myers [1995a]}). In the case of a non-minimally
coupled scalar field, naively it appears that the divergences appearing
in the statistical mechanical entropy will be different from those
in the renormalization of Newton's constant. However, one
obtains the correct renormalization  after taking into account
the appropriate degrees of freedom at the horizon
(\cite{Demers, Lafrance, Myers [1995b]}).

\enddocument